\documentclass[11pt,a4paper]{article}
\usepackage{enumerate,amsmath,amsthm,mathrsfs,array,pstricks, amssymb}
\allowdisplaybreaks[1]

\newtheoremstyle{bthm}{\baselineskip}{\baselineskip}{\slshape}{}{\bfseries}{:}{\newline }{}
\newtheoremstyle{bex}{\baselineskip}{\baselineskip}{}{}{\sffamily}{:}{\newline }{}
\theoremstyle{bthm}
\newtheorem{thm}{Theorem}

\newtheorem{rem}{Remark}
\begin{document}
\title{A sharp lower bound for the Wiener index of  a graph}
\author{R. Balakrishnan\\ Srinivasa Ramanujan Centre\\Kumbakonam--612 001, India\\ 
email: mathbala@sify.com 
\and
N. Sridharan\\
Department of Mathematics\\
Alagappa University\\
Karaikudi--630 003, India\\
email: math\_sridhar@yahoo.co.in
\and  K. Viswanathan Iyer\\
Department of Computer Science and Engg.\\ National Institute of Technology\\ 
Trichy--620 015, India\\
email: kvi@nitt.edu
}
\date{}
\maketitle
\renewcommand{\baselinestretch}{1.58}\small\normalsize
\begin{abstract}
Given a simple connected undirected graph $G$, the Wiener index $W(G)$ of $G$ is defined as half the sum of the distances over all pairs of vertices of $G$.  In practice, $G$ corresponds to what is known as the \emph{molecular graph} of an  organic compound.   We obtain a sharp lower bound for $W(G)$ of an arbitrary graph  in terms of the order, size and  diameter of $G$.  
\end{abstract}
\section{Introduction}
Let $G=\big(V(G),\,E(G)\big)$ be a simple  connected undirected graph of order $n$ and size $m$. Given any two vertices $u,\,v$ of $G$, let  $d(u,\, v)$ denote the distance between $u$ and $v$. 
The Wiener index $W(G)$ of the graph $G$ is defined by
\[ W(G)=\frac{1}{2}\sum_{u, v\in V(G)}d(u,\, v),\]
where the summation is over all possible pairs $u,\, v \in V(G)$.  Our notation and terminology are as in~\cite{bala_ranga}.

Wiener index, first proposed in \cite{wiener}, is currently a widely used topological index (see, for example, \cite{xu}) and has applications in modern drug design~\cite{es2001}.  For more information about Wiener index in chemistry and in mathematics, see~\cite{polansky}, \cite{dobry1} and \cite{dobry2}. In this paper, for any graph $G$, we obtain a sharp lower bound for $W(G)$ in terms of the order, size and diameter of $G$.

\section{Lower bound for $W(G)$}
Let $S_2(G)$ be the set of all 2-subsets of $V(G)$ (that is, the set of all unordered pairs of distinct vertices of $G$).  We can then equivalently define $W(G)$ as
\begin{equation}
W(G)=\sum_{\{u,v\}\in S_2(G)}d(u,v).\label{bound}
\end{equation}
Let $P=u_0u_1\ldots u_d$ be a diametral path of $G$, so that $d(u_0,u_d)=d$, the diameter of $G$.  We partition $S_2(G)$ into disjoint sets $X,Y$ and $Z$ defined as follows:
\begin{align*}
X&=\{\{u,v\}\in S_2(G)\vert \quad \text{both}\  u,v\in P\},\\
Y&=\{\{u,v\}\in S_2(G)\vert \quad \text{none of $u$ and $v$ is in $P$}\},\quad\text{and}\\
Z&=\{\{u,v\}\in S_2(G)\vert \quad \text{one of $u$ and $v$ alone is in $P$}\}.
\end{align*}
It follows that
\[
\vert X\vert =\frac{d(d+1)}{2};\quad \vert Y\vert =\frac{(n-d-1)(n-d-2)}{2};\quad \vert Z\vert =(n-d-1)(d+1).
\]
From~\eqref{bound}, we have
\begin{align}
W(G)&=\sum_{\{u,v\}\in S_2(G)}\big(2+(d(u,v)-2)\big)\nonumber\\
&=\sum_{\{u,v\}\in S_2(G)}2\; + \sum_{\{u,v\}\in S_2(G)}\big(d(u,v)-2\big)\label{wpapeqn12}\\
&= n(n-1) 
+\sum\limits_{\substack{\{u,\,v\}\in S_2 (G)\\d(u,\,v)=1}} \big(d(u,\,v)-2\big) + \mathop{\sum}\limits_{\substack{\{u,\,v\}\in S_2 (G)\\d(u,\,v)\geq 2}} \big(d(u,\,v)-2\big)\nonumber\\
\intertext{(since $\vert S_2(G)\vert  = n(n-1)/ 2)$}
&= \big(n(n-1) -m\big) + \mathop{\sum}\limits_{\substack{\{u,\,v\}\in S_2(G)\\d(u,\,v)\geq 2}} \big(d(u,\,v)-2\big)\nonumber\\
&\ge (n(n-1) - m) + \sum_{\substack{\{u,\,v\} \in X\cup Z\\d(u,\,v)\ge2}} \big(d(u,\,v)-2\big)\label{wpapeqn13}.
\end{align}
For $0\le k \le(d-1)$ in $X$, there are $(d-k)$ pairs $\{u,\,v\}$ with $d(u,\,v) = 1 + k$.  Hence
\begin{align}
\sum_{\substack{\{u,\,v\} \in X\\d(u,\,v)\ge 2}} \big(d(u,\,v) - 2\big) &= (d-2)1 + (d-3)2 + \cdots + 1(d-2)\nonumber\\
							 &= \frac{d(d-1)(d-2)}{6}. 	
\end{align}
We next obtain a lower bound for the summation term on the right hand side of equation~\eqref{wpapeqn13}.  We first assume that $d\ge 5$.  Fix one vertex $w$ in $V(G)\setminus V(P) $, where $V(P)$ is the set of vertices of $P$.  Then, by triangle inequality, we have
\begin{align}
d(u_i,\,w) + d(w,\,u_{d-i})\ge \; &d(u_i,\,u_{d-i}) = d-2i,\\ 
&\text{for}\quad 0\le i < (d-3)/2.\nonumber
\end{align}
Therefore, for each of the $(n-d-1)$ choices of $w$, we have
\begin{align}
\sum_{d(u_i,w) \ge 2} \big(d(u_i,\,w)-2\big) \nonumber \\
\quad \ge \sum_{i=0}^d \big(d(u_i,\,w)-2\big)
&\ge \sum_{i=o}^{\lfloor\frac{d-3}{2}\rfloor} (d(u_i,\,w) + d(u_{d-i},\,w)-4)\label{wpapeqn16}\\
&\ge \sum_{i=o}^{\lfloor\frac{d-3}{2}\rfloor} (d-2i-4).\label{wpapeqn17}
\end{align} 
Since each summand on the right side of~\eqref{wpapeqn16} is nonnegative,  so is each summand on the right side of~\eqref{wpapeqn17}.  Hence the term on the right side of~\eqref{wpapeqn16} is
\begin{align*}
& \ge 
\begin{cases}
(d-4)+(d-6) +\cdots + 5+3+1 \quad \text{if $d$ is odd, and}\\
(d-4) + (d-6) + \cdots + 2 \quad \text{if $d$ is even}.
\end{cases}\\[2ex]
&= 
\begin{cases}
(\frac{d-3}{2})^2 \quad \text{if $d$ is odd, and}\\
\frac{(d-2)(d-4)}{4}\quad \text{if $d$ is even}.
\end{cases}
\end{align*}
Thus, for each fixed $w \in V(G) \setminus V(P)$ we have 
\[
\sum_{(u,\,w)\in Z} (d(u,\,w)-2) = 
\begin{cases}
(\frac{d-3}{2})^2 \quad \text{if} \; d \;\text{is odd, and}\\
\frac{(d-2)(d-4)}{4}\quad\text{if} \; d\;\text{is even}. 
\end{cases}
\]
In conclusion, for $d \ge 5$, we have
\[
W(G) \geq 
\begin{cases}
(n(n-1)-m + \frac{d(d-1)(d-2)}{6} + \frac{(n-d-1)(d-3)^2}{4},\quad \text{if $n$ is odd}, \\
(n(n-1)-m) + \frac{d(d-1)(d-2)}{2} + \frac{(n-d-1)(d-2)(d-4)}{4}, \; \text{if $n$ is even}. 
\end{cases}
\]
We now consider the cases when $d = 2,\,3\,$ and 4.
If $d=2$, $d(u,\,v) = $ 1 or 2. Hence from~\eqref{wpapeqn12}, we get 
\begin{align}
W(G) &= n(n-1) + \sum_{d(u,\,v)=1} (1-2)\nonumber\\
		 &= n(n-1)-m.
\end{align}
Now consider the case when $d = 3$. Then $d(u,\,v) = $1,\,2 or 3.  Hence from~\eqref{wpapeqn12} we get
\begin{align*}
W(G) &= n(n-1)-m + \sum_{d(u,\,v)=3} (d(u,\,v)-2)\\
&\ge n(n-1)-m + 1,
\end{align*}
as there is at least one pair with $d(u,\,v)=3$.  If $d = 3$, $\dfrac{d(d-1)(d-2)}{6} = 1$.  Hence we have 
\[
W(G) \ge n(n-1)-m + \frac{d(d-1)(d-2)}{6}.
\]
Finally if $d = 4$, we have 
\begin{align*}
W(G) \ge n(n-1)-m  &+ \sum_{d(u,\,v) =3} (d(u,\,v)-2) \\ &+ \sum_{d(u,\,v)=4} (d(u,\,v)-2)\\
&\geq\;n(n-1)-m+2+2\tag*{(since $d(u_0,\,u_d)=d= 4$, $d(u_0,\,u_3)=2=d(u_1,\,u_4)$)}.
\end{align*}
Hence
\begin{align*}
W(G)&\geq n(n-1)-m+\frac{d(d-1)(d-2)}{6} \tag*{\big(\text{as $\frac{d(d-1)(d-2)}{6}=4$ in this case.\big)}}.
\end{align*}
To conclude, we have proved the following result:
\begin{thm}
\label{sharp-thm-1}
If $G$ is any graph of order $n$, size $m$  and diameter $d\ge 2$ then
\begin{align}
W(G)\geq
\begin{cases}
n(n-1)-m+\frac{d(d-1)(d-2)}{6}+\frac{(n-d-1)(d-3)^2}{4},&\text{if $n$ is odd.}\\
n(n-1)-m+\frac{d(d-1)(d-2)}{6}+\frac{(n-d-1)(d-2)(d-4)}{4},&\text{if $n$ is even.}\label{new-9}
\end{cases}
\end{align}
\end{thm}
\begin{rem}
Let the maximum degree of $G$ be $\Delta$.  The Moore bound (see for example~\cite{ms05}) gives an upper bound for $n$ in terms of $\Delta$ and the diameter $d$:
\begin{equation}
n\leq 
\begin{cases}
1+ \Delta \frac{(\Delta -1)^d-1}{\Delta -2}& \text{if $\Delta >2.$}\\
2d+1 & \text{if $\Delta=2.$}
\end{cases}\label{new-10}
\end{equation}
\noindent
From~\eqref{new-10}, a lower bound for $d$ in terms of $n$ and $\Delta$ can be obtained and if this is used in~\eqref{new-9}, a lower bound for $W(G)$ can be obtained in terms of $n$, $m$ and $\Delta$.
\end{rem}
\begin{rem}
If $G$ is known to be planar and if $d$ is known to be bounded by a constant then we can compute the exact value of $d$ in time $O(n)$ as shown in \cite{epp99}.  In turn, this implies that the lower bound on $W(G)$ in~\eqref{new-9} can be computed in time $O(n)$ if $G$ is planar.
\end{rem} 
We observe that the lower bound given in Theorem~\ref{sharp-thm-1} is sharp.  The graphs $P_n$, $K_{1,m}$ ($m$:odd), $C_3\, \Box\, K_2$ and the Petersen graph attain the bound.

\end{document}